\documentclass{css2016}
\usepackage{amssymb,amsmath}
\usepackage{epsfig}

\newcommand{\PACS}{\MSC}

\title{Local Hubble Expansion: \\
       Current State of the Problem}

\author{Yurii V. Dumin \\
\vskip 2mm {\small
$^1$P.\,K.~Sternberg Astronomical Institute
of M.\,V.~Lomonosov Moscow State University \\
Universitetskii prosp. 13, 119992, Moscow, Russia \\
$^2$Space Research Institute of the Russian Academy of Sciences \\
Profsoyuznaya str.\ 84/32, 117997, Moscow, Russia \\
dumin@yahoo.com, dumin@sai.msu.ru
}}

\abstract{
We present a brief qualitative overview of the current state
of the problem of Hubble expansion at the sufficiently small
scales (e.g.,~in planetary systems or local intergalactic volume).
The crucial drawbacks of the available theoretical treatments are
emphasized, and the possible ways to avoid them are outlined.
Attention is drawn to a number of observable astronomical phenomena
that could be naturally explained by the local Hubble expansion.
}



\begin{document}

\maketitle
\setcounter{page}{23}

\section{Introduction: theoretical approaches to the problem of
         local Hubble expansion}
\label{sec:Introduction}

The problem of small-scale cosmological effects has a long history:
the question if planetary systems are affected by the universal
Hubble expansion was posed by McVittie as early as 1933~\cite{mvi33},
i.e., approximately at the same time when the concept of Hubble expansion
became the dominant paradigm in cosmology.
Although this question never was a hot topic, the corresponding papers
occasionally appeared in the astronomical literature in the subsequent
eight decades~\cite{and95,coo98,dav03,dic64,dom01,ein45,far07,kli05,lin05,%
noe71,ser07}.
Using quite different physical models and mathematical approaches,
most of these authors arrived at the negative conclusions.
As a result, it is commonly believed now%
\footnote{
One of a few exceptions is a~short review by Bonnor~\cite{bon00},
which appealed for a~critical reconsideration of the available studies.
}
that Hubble expansion should be
strongly suppressed or absent at all at the sufficiently small scales,
for example, in planetary systems or inside galaxies.

However, a surprising thing is that the commonly-used arguments
not only prohibit the local Hubble expansion but also strongly contradict
each other.
For example, the most popular criterion for the suppression of
Hubble expansion (especially, among the observational astronomers)
is just a gravitational binding of the system, e.g., determined by
the virial theorem of classical mechanics~\cite{lan76}.
Namely, if mass of the particles concentrated in the system becomes
so large that the corresponding energy of gravitational interaction
approaches by absolute value the double kinetic energy, then orbits of
the particles should be bounded, i.e., no overall expansion of\, the\, system\,
is\, possible.\,
In\, other\, words,\, just\, the\, classical\, forces\, of gravitational attraction\,
break\, the\, global\, Hubble\, flow\, in\, the\, regions\, of\, local\, mass\, enhancement.

\begin{figure}[t]
\begin{center}
\includegraphics[width=7cm]{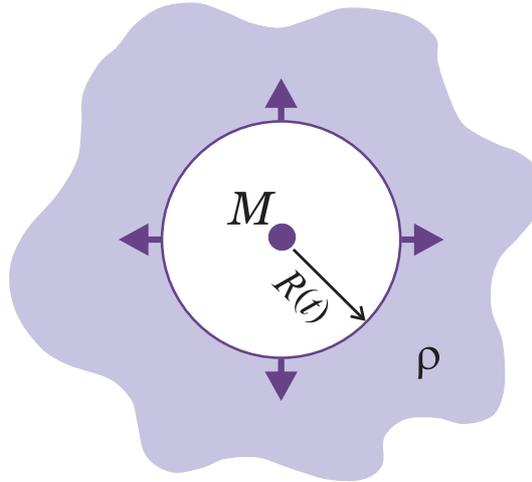}

\medskip

\caption{\label{fig:E-S_th_2}
Schematic illustration of Einstein--Straus theorem.}
\end{center}
\end{figure}

On the other hand, yet another well-known theoretical argument against
the local Hubble expansion, based on the self-consistent theoretical
analysis in the framework of General Relativity~(GR), is the so-called
Einstein--Straus theorem~\cite{ein45}, illustrated in
Figure~\ref{fig:E-S_th_2}:
Let us consider a uniform distribution of the background matter with
density~$ \rho $ and then assume that substance in a spherical volume
with radius~$ R $  is cut off and concentrated in its center,
thereby forming the point-like mass~\hbox{$ M \! = ( 4 \pi / 3 ) R^3 \rho $}.
Then, according to the this theorem, there will be no Hubble expansion
inside the empty cavity, but the Hubble flow is restored again beyond
its boundary with the background matter distribution (and this boundary
itself moves exactly with Hubble velocity).

It is important to emphasize that, as distinct from the first criterion,
there is no any excessive mass in the above-mentioned sphere and, moreover,
the Hubble expansion is absent just in the empty space rather than in
the region of mass enhancement.
In\, principle,\, this\, fact\, is\, quite\, natural:\, according\, to\, the\, standard\,
GR\, formula, Hubble constant~$ H $ is related to the local energy
density~$ \rho $ in the spatially-flat Universe as%
\footnote{
We use everywhere the system of units where the speed of light is equal
to unity ($ c \equiv 1 $) and, therefore, there is no difference between
the mass- and energy-density.
}
\begin{equation}
\label{eq:Hubble_const}
H = \, \sqrt{\frac{8 \pi G}{3} \, \rho } \: ,
\end{equation}
where $ G $~is the gravitational constant.
So, from the relativistic point of view, it is not surprising that Hubble
constant tends to zero when the energy density disappears%
\footnote{
Of course, strictly speaking, formula~(\ref{eq:Hubble_const}) is applicable
only to the totally uniform Universe.
}.

Therefore, the above discussion demonstrates that the attempts to treat
the problem of local Hubble expansion in terms of the classical
gravitational forces can be very misleading.
Indeed, the global Hubble expansion exists even in the perfectly-uniform
Universe, where there are no any ``classical'' gravitational forces at all
(since such forces can be produced only by nonhomogeneity of mass
distribution).
In other words, it should be kept in mind that Hubble expansion corresponds
to another ``degree of freedom'' of the relativistic gravitational field
as compared to the degrees of freedom reduced to the classical
gravitational forces.

Unfortunately, a lot of textbooks tried to estimate the local Hubble
expansion in terms of the ``classical'' gravity or just \textit{postulated}
its absence in the small-scale systems.
A typical example is the famous textbook~\cite{mis73}, where
the behavior of small-scale systems (galaxies) in the globally-expanding
Universe was pictorially described as a set of coins pinned to
the surface of an inflating ball (see Figure~27.2 in the above-cited
book), but no justification for such a picture was given.

\section{Hubble expansion in the dark-energy-dominated cosmology}
\label{sec:Dark_energy}

Because of the oversimplified geometry of the Einstein--Straus model
(particularly, a~presence of the void, which can hardly have a reasonable
astrophysical interpretation), it is desirable to consider not so
idealized situations.
Unfortunately, a serious obstacle in this way is the problem of separation
between the peculiar and Hubble flows of matter in a spatially
inhomogeneous system.
Namely, if there is no empty cavity, and boundary with the background
matter distribution is not perfectly sharp, as in
Figure~\ref{fig:E-S_th_2}, then substance in the vicinity of the central
mass will experience a~quite complex radial motion in the course of time,
depending on the initial conditions.
In general, we do not have any universal criterion to answer the question:
what part of this motion should be attributed to the Hubble flow?

Fortunately, the situation is simplified very much in the case of
idealized dark-energy-dominated Universe, where the entire
cosmological contribution to the energy--momentum tensor of GR~equations
is produced by the $ \Lambda $-term (cosmological constant).
The $ \Lambda $-term is distributed, by definition, perfectly uniform
in space and, therefore, does not experience any back reaction from
the additional (e.g.,~point-like) mass%
\footnote{
We do not discuss here the models with ``dynamical'' dark energy
(where $ \Lambda $-term is replaced by a new field), because they are not
so necessary to explain the available observational data.
}.
As a~result, it becomes not so difficult to consider the ``restricted
cosmological two-body problem'', i.e., motion of a test particle
in the local gravitational field of the central mass embedded
into the cosmological background formed by the $ \Lambda $-term%
\footnote{
According to the standard terminology of celestial mechanics,
the term ``restricted'' implies that one of the bodies (test particle)
has infinitely small mass.
}.
From our point of view, just this model enables one to get the simplest
but reliable estimate for the magnitude of the local Hubble expansion,
which can be used as a~benchmark in more sophisticated studies.

The above-mentioned problem can be separated into two steps:
Firstly, using GR~equations, one needs to find a space--time metric
of the point-like mass~$ M $ against the $ \Lambda $-background.
Secondly, using the standard geodesic equations, we should calculate
the trajectories of test particles in this metric.

The first task was actually solved long time ago, in~1918, by
Kottler~\cite{kot18}.
The required metric reads as%
\footnote{
It is often called in the modern literature
the Schwarzschild--de~Sitter metric; although, from our point of view,
this term is not sufficiently correct.
}

\begin{eqnarray}
\label{eq:Kottler_metric}
{\rm d} s^2 \! & \!\! = \! & -\, {\biggl( 1 - \frac{2 G M}{r}
  - \frac{\Lambda r^2}{3} \biggr)}\, {\rm d}{t}^2
\\[2mm]
\nonumber
  & & + \, {\biggl( 1 - \frac{2 G M}{r}
  - \frac{\Lambda r^2}{3} \biggr)}^{\!\! -1} \!\! {\rm d}r^2 \!
  + \: r^2 ( {\rm d}{\theta}^2 \!
  + \, {\sin}^2{\theta} \, {\rm d}{\varphi}^2 ) \: ,
\end{eqnarray}

\smallskip

\noindent for more general discussion, see also~\cite{kra80}.

Just this metric was widely used starting from the early 2000's~--- when
the importance of $ \Lambda $-term in cosmology was clearly
recognized~--- to study the motion of test particles.
The quite sophisticated mathematical treatments can be found, for example,
in papers~\cite{bal06,hac08}; and the respective formulas were used
for the analysis of observational data on planetary dynamics in the Solar
system~\cite{car98,ior06,kag06}.
Unfortunately, the original Kottler metric~(\ref{eq:Kottler_metric})
does not possess the correct cosmological asymptotics at infinity
(which is not surprising, since it was derived well before a~birth of
the modern cosmology).
The above-cited works, of course, reveal some features of
particle dynamics in the dark-energy-dominated Universe, but they are
unrelated (or, probably, partially related) to the Hubble expansion
by itself.

So, to study effects of the Hubble expansion \textit{per se},
it is necessary to transform metric~(\ref{eq:Kottler_metric}) to
the standard Robertson--Walker coordinates, commonly used
in the cosmological calculations.
Such a~procedure was performed in our paper~\cite{dum07};
and the resulting expressions for the ``cosmological'' Kottler metric
can be found there.
Next, this metric should be used to solve the geodesic equations for
a~test particle moving in the field of the central mass~\cite{dum11}:

\begin{eqnarray}
\label{eq:Geodesic_1}
&& \!\!\!\!\!\!\!\!\!\!\!\!
2 \: \bigg[ 1 - \frac{r_g}{r} \bigg( 1 - \frac{t}{r_{\Lambda}} \bigg) \bigg] \:
  \ddot{t} \:
- \: 4 \, \frac{r_g}{r_{\Lambda}} \, \ddot{r} \:
+ \: \frac{r_g}{r_{\Lambda}} \, \frac{1}{r} \: {\dot{t}}^{\, 2} \:
+ \: 2 \, \frac{r_g}{r^2} \bigg( 1 - \frac{t}{r_{\Lambda}} \bigg) \, \dot{t} \,
  \dot{r} \:
\\
\nonumber
&& \qquad
+ \: \frac{1}{r_{\Lambda}} \, \bigg( 2 + \frac{r_g}{r} \bigg) \:
  {\dot{r}}^{\, 2}
+ \: 2 \, \frac{r^2}{r_{\Lambda}} \, {\dot{\varphi}}^{\, 2} \:
= \: 0 \; ,
\\[3mm]
\label{eq:Geodesic_2}
&& \!\!\!\!\!\!\!\!\!\!\!\!
4 \, \frac{r_g}{r_{\Lambda}} \, \ddot{t} \:
+ \: 2 \, \bigg[ 1 + 2 \, \frac{t}{r_{\Lambda}}
  + \frac{r_g}{r} \bigg( 1 + \frac{t}{r_{\Lambda}} \bigg) \bigg] \, \ddot{r} \:
+ \: \frac{r_g}{r^2} \bigg( 1 - \frac{t}{r_{\Lambda}} \bigg) \,
  {\dot{t}}^{\, 2} \:
\\
\nonumber
&& \qquad
+ \: \frac{2}{r_{\Lambda}} \bigg( 2 + \frac{r_g}{r} \bigg) \, \dot{t} \,
  \dot{r} \:
- \: \frac{r_g}{r^2} \bigg( 1 + \frac{t}{r_{\Lambda}} \bigg) \,
  {\dot{r}}^{\, 2}
- \, 2 \, r \bigg( 1 + 2 \, \frac{t}{r_{\Lambda}} \bigg) \,
  {\dot{\varphi}}^{\, 2} \:
= \: 0 \; ,
\\[3mm]
\label{eq:Geodesic_3}
&& \!\!\!\!\!\!\!\!\!\!\!\!
r \bigg( 1 + 2 \, \frac{t}{r_{\Lambda}} \bigg) \, \ddot{\varphi} \:
+ \: 2 \, \frac{r}{r_{\Lambda}} \: \dot{t} \dot{\varphi} \:
+ \: 2 \, \bigg( 1 + 2 \, \frac{t}{r_{\Lambda}} \bigg) \, \dot{r}
  \dot{\varphi} \:
= \: 0 \; .
\end{eqnarray}
Here, as distinct from formula~(\ref{eq:Kottler_metric}),
$ t $~and~$ r $~are the Robertson--Walker coordinates; and
dot denotes a derivative with respect to the proper time of
the moving particle.

\begin{figure}[t]
\begin{center}
\includegraphics[width=\textwidth]{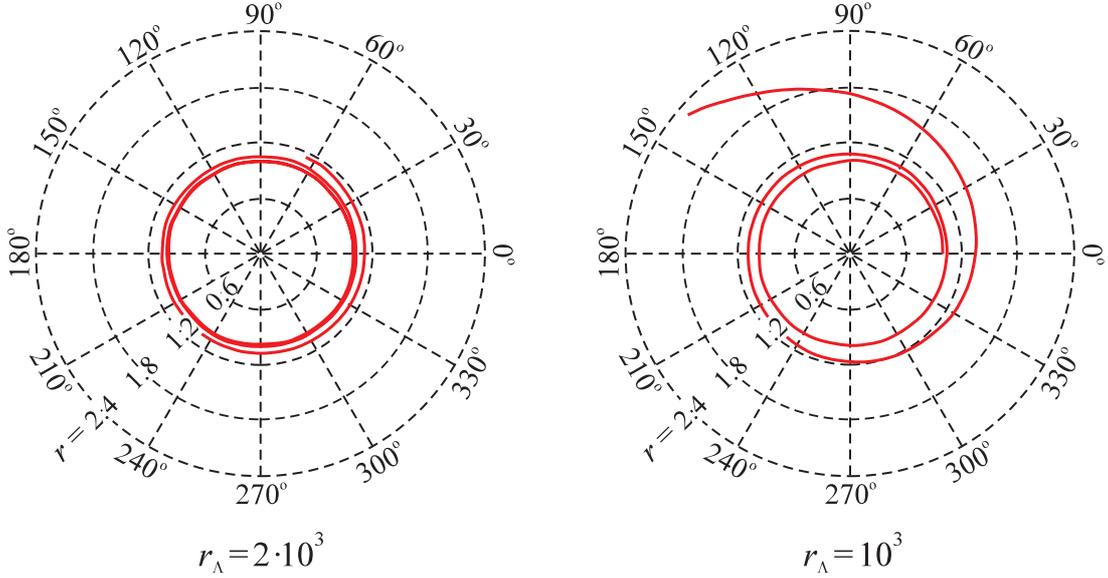}
\caption{\label{fig:orbits}
Orbits of a test body in the field of the central mass at $ r_g = 10^{-2} $
and various values of $ r_{ \Lambda } $, assuming that $ R_0 \! = 1 $.}
\end{center}
\end{figure}

An important feature of these equations is that they involve three
characteristic spatial scales --- Schwarzschild radius $ r_g  = 2  G M $,
de~Sitter radius $ r_{ \Lambda }  = \sqrt{ 3 / \Lambda } $, and
the initial radius of orbit of the test body
(e.g., a planet)~$ R_0 $ --- which differ from each other by many
orders of magnitude.
For example, in the case of the Earth--Moon system (where Earth is
the central mass; and Moon, the test body), we have:
$ r_g{\sim}10^{-2} $\,m,
$ R_0{\sim}10^9 $\,m, and
$ r_{ \Lambda }{\sim}10^{27} $\,m.
This makes the problem of accurate numerical integration very hard.

However, for simplicity --- just to reveal the possibility of local Hubble
expansion in the gravitationally-bound system --- we can consider a~toy model,
where these parameters differ from each other not so much, e.g., by only
two or three orders of magnitude.
For example, let us take the initial orbital radius as the unit of length
(i.e., $ R_0 \equiv 1 $);
and let the Schwarzschild radius be $ r_g  = 10^{-2} $, and
de~Sitter radius $ r_{ \Lambda }  = 10^3 $ or~$ 2{\cdot}10^3 $.
The corresponding numerical orbits are presented in Figure~\ref{fig:orbits}.
As is seen, when $ \Lambda $ (i.e., the dark-energy density) increases
and, respectively, $ r_{ \Lambda } $~decreases, the orbits become
more and more spiral.
In other words, \textit{a~test particle orbiting about the central mass
can really experience the local Hubble expansion.}
This quantitative analysis argues against the commonly-accepted intuitive
point of view that the remote cosmological action would result just in
a partial compensation of the gravitational attraction to
the center, i.e., the orbit will be slightly disturbed but remain
stationary~\cite{lin05}.
According to our calculations, the secular (time-dependent) effects are
really possible.

Unfortunately, it is not so easy to get the reliable numerical values of
such an effect in the realistic planetary systems, because of
the above-mentioned huge difference in the characteristic scales and
the need for integration over a very long time interval.
Besides, since the set of equations is strongly nonlinear, it is difficult
to predict how the other kinds of celestial perturbations (e.g., by
the additional planets) will interfere with the secular Hubble-type effects.
Moreover, it is unclear in advance if the local Hubble expansion will
follow the standard linear relation:
\begin{equation}
\label{eq:Hubble_local_rate}
\dot r = H_0^{\rm (loc)} \, r \, ,
\end{equation}
where $ H_0^{\rm (loc)} $~is the local Hubble constant (which, generally
speaking, can be different from the global one).
In principle, the corresponding relation in the vicinity of
the central massive body might be substantially nonlinear.
So, all these questions are still to be answered.

\bigskip

\section{Observable footprints of the local Hubble expansion}
\label{sec:Observations}

\smallskip

A crucial factor supporting the interest to a probable manifestation
of Hubble expansion at the small scales is that there is a number of
observable phenomena --- both in the Solar system and local intergalactic
volume --- that could be naturally explained by the local Hubble expansion.
A detailed list of such effects in the Solar system can be found in
papers~\cite{kri12,kri15}.
Particularly, they are:
\begin{itemize}
\item
the so-called faint young Sun problem (i.e., the insufficient luminosity
of the young Sun to support development of the geological and biological
evolution on the Earth),
\item
the problem of liquid water on Mars (which actually has the same origin as
the above-mentioned one),
\item
the anomalous rate of recession of the Moon from the Earth (called also
the lunar tidal catastrophe),
\item
the long-term dynamics of the so-called fast satellites of Mars,
Jupiter, Uranus, and Neptune,
\item
the efficiency of formation of Neptune and comets in the Kuiper belt
from the protoplanetary disk.
\end{itemize}

\medskip

\subsection{The lunar tidal catastrophe}
\label{subsec:Moon}

\begin{figure}[t]
\begin{center}
\includegraphics[width=9.5cm]{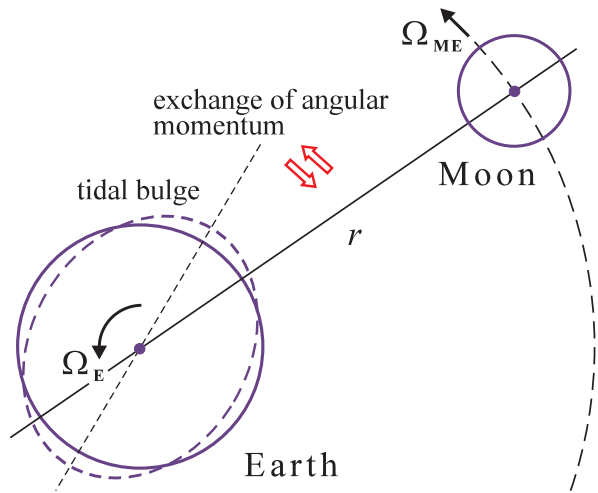}
\caption{\label{fig:tidal_inter}
Sketch of the tidal interaction between the Earth and Moon.}
\end{center}
\end{figure}

From our point of view, the most appealing example for the existence of
the local Hubble expansion is the anomalous Earth--Moon recession rate.
Namely, it was known for a long time that tidal interaction results in
the deceleration of \textit{proper} rotation of
the Earth~$ {\Omega}_{\rm E} $ and acceleration of \textit{orbital} rotation
of the Moon~$ {\Omega}_{\rm ME} $~\cite{kau68}.
This is pictorially explained in Figure~\ref{fig:tidal_inter}:
since $ {\Omega}_{\rm E}  >  {\Omega}_{\rm ME} $, a~tidal bulge on
the Earth's surface is slightly shifted forward (in the direction of
Earth's rotation) because of the finite-time relaxation effects.
Such a shifted bulge pulls the Moon forward, thereby accelerating it;
and simultaneously, due to the back reaction, the proper rotation of
the Earth decelerates.

The increasing orbital momentum of the Moon results in the increase of
its distance from the Earth with the following rate:
\begin{equation}
\label{eq:tid_inter}
\dot r = k \, {\dot T}_{\rm E} \, ,
\end{equation}
where $T_{\rm E}$~is the Earth's diurnal period, and
$ k  =  1.81{\cdot}10^5 $\,cm/s~\cite{dum03}.
So, if secular variation in the length of day is known from astrometric
observations, relation~(\ref{eq:tid_inter}) can be used to derive the rate
of secular increase in the lunar orbit~$ \dot r $.

\begin{table}[t]
\begin{center}
\begin{tabular}{@{}lll@{}}
\hline\\
&&\\[-20pt]
\multicolumn{1}{c}{Method} &
\multicolumn{1}{c}{Measurement by the lunar} &
\multicolumn{1}{c}{Estimate from the Earth's} \\
& \multicolumn{1}{c}{laser ranging} &
\multicolumn{1}{c}{tidal deceleration}
\\[6pt]
\hline\\
&&\\[-18pt]
\hphantom{i} Effects involved &
(1) geophysical tides &
(1) geophysical tides \\
& (2) local Hubble expansion & \\
&&\\[-7pt]
\hphantom{i} Numerical value &
$3.8{\pm}0.1$~cm/yr &
$1.6{\pm}0.2$~cm/yr
\\[8pt]
\hline
\end{tabular}
\end{center}
\caption{\label{tab:Moon-Earth}
Relative contributions of various processes to
the recession rate of the Moon from the Earth.}
\end{table}

On the other hand, the same quantity can be measured immediately by
the lunar laser ranging~(LLR). This became possible since the early~1970's,
when a~few optical retroreflectors were installed on the lunar surface.
The accuracy of LLR quickly improved in the subsequent two decades,
and its errors were reduced to 2--3\,cm, which enabled ones
to measure immediately the secular expansion of the lunar orbit~\cite{dic94}.
Surprisingly, the measured value of~$ \dot r $ turned out to be
substantially greater than the value obtained from
formula~(\ref{eq:tid_inter}), as summarized in
Table~\ref{tab:Moon-Earth}~\cite{dum08}.

Then, a~lot of attempts were undertaken to reduce this discrepancy.
Namely, the value presented in the last column of the table corresponds to
\begin{equation}
\label{eq:T_E_dot}
{\dot T}_{\rm E}  =  (8.77{\pm}1.04){\cdot}10^{-6}{\,}{\rm s/yr} \, .
\end{equation}
It was derived from a series of astronomical observations accumulated
since the middle of the 17th~century, when telescopic data became available
(they are compiled, for example, in monograph~\cite{sid02}).
In principle, the period of three centuries might be insufficiently long,
because the length of day~$ T_{\rm E} $ experience also some quasi-periodic
variations on the longer time scales, which can affect the linear
trend~(\ref{eq:T_E_dot}).

One of the ways to get around this obstacle is to employ the ancient data
on eclipses, which cover the period over two millennia.
Such an approach was pursued by a number of researchers (e.g.,
review~\cite{ste84}), and the obtained values of~$ {\dot T}_{\rm E} $
sometimes enabled them to get a reasonable agreement with LLR data.
However, the various sets of ancient observations give the results
different from each other by almost two times, and it is not clear
\textit{a priori} which of them are more reliable.

Yet another idea to avoid the discrepancy presented in
Table~\ref{tab:Moon-Earth} is to take into account a secular variation in
the Earth's moment of inertia, which is commonly characterized by
the second gravitational harmonic coefficient~$ \dot J_2 $.
Its decreasing trend at the present time is assumed to be caused
by the so-called viscous rebound of the solid Earth from the decrease in
load due to the last deglaciation. (Namely, the Earth was compressed by
the ice caps in polar regions during the glacial period and now restores
its shape.)
The first determination of the above-mentioned parameter by Lageos
satellite~\cite{yod83} led to the value
$ \dot J_2  = -3{\cdot}10^{-11}/{\rm yr} $, which seemed to be consistent
with LLR~data.
However, as was established later, such a determination may be very
unreliable~\cite{bou04} and even can give the opposite sign
of~$ \dot J_2 $~\cite{cox02}.

In view of the above difficulties, a~promising explanation of the discrepancy
$ 2.2{\pm}0.3$\,cm/yr in Table~\ref{tab:Moon-Earth} can be based just on
the presence of local Hubble expansion.
Assuming validity of the standard relation~(\ref{eq:Hubble_local_rate}),
this corresponds to the value of the local Hubble constant
\begin{equation}
\label{eq:H_0_loc}
H_0^{\rm (loc)} = \, 56{\pm}8~{\rm (km/s)/Mpc} \, ,
\end{equation}
which is quite close to its ``global'' value~$ H_0 $.
So, such an interpretation is not meaningless.

Unfortunately, as was mentioned in Section~\ref{sec:Dark_energy},
by now we cannot reliably explain this quantity in terms of
parameters of the Earth--Moon system because of the problems in
the numerical integration of the equations of motion.
Instead, we shall present here a~more crude but universal estimate
of the relation between the local and global Hubble rates, which is
actually applicable to any ``small-scale'' system.

It is reasonable to assume that the local Hubble expansion is formed
only by the uniformly-distributed dark energy ($ \Lambda $-term), while
the irregularly distributed (clumped) forms of matter affect the rate
of cosmological expansion only at the sufficiently large distances,
where they can be characterized by their average values.
(At~smaller distances, the clumped forms of matter manifest themselves
by the ``classical'' gravitational forces.)
So, if the Universe is spatially flat and filled only with dark energy
and the dust-like matter with densities~$ \rho_{{\Lambda}0} $ and~$\rho_{{\rm D}0}$, respectively,
then general expression~(\ref{eq:Hubble_const}) can be rewritten as%
\footnote{
Subscripts ``0'' denote here the values of the corresponding quantities
at the present time.
}
\begin{align}
\label{eq:H_0-rho_glob}
H_0  & =  \sqrt{\frac{8 \pi G}{3}}
      \; \sqrt{\, \rho_{{\Lambda}0} + \rho_{{\rm D}0}} \; ,
\\[3mm]
\label{eq:H_0-rho_loc}
H_0^{\rm (loc)}  & =   \sqrt{\frac{8 \pi G}{3}}
      \; \sqrt{\, \rho_{{\Lambda}0}} \; .
\end{align}
Therefore, a ratio of the local to global Hubble constants will be
\begin{equation}
\label{eq:H_0_ratio}
\frac{H_0^{\rm (loc)}}{H_0} =
{\Bigg[ 1 +
  \frac{\Omega_{{\rm D}0}}{\Omega_{{\Lambda}0}} \, \Bigg]}^{-1/2} \! ,
\end{equation}
where $ {\Omega}_{{\Lambda}0} \! = {\rho}_{{\Lambda}0} / {\rho}_{\rm cr} $
and $ {\Omega}_{{\rm D}0} \! = {\rho}_{{\rm D}0} / {\rho}_{\rm cr} $
are the corresponding relative densities.

\pagebreak

Taking for a crude estimate $ \Omega_{{\Lambda}0}  =  0.75 $ and
$ \Omega_{{\rm D}0}  =  0.25 $, we arrive at
\begin{equation}
\label{eq:H_0_ratio_num}
H_0 / H_0^{\rm (loc)} \approx \, 1.15 \, .
\end{equation}
Consequently, the local value~(\ref{eq:H_0_loc}) corresponds to
the global value
\begin{equation}
\label{eq:H_0_glob}
H_0 = \, 65{\pm}9~~{\rm (km/s)/Mpc} \, ,
\end{equation}
which is in a good agreement with the modern cosmological data
(especially, based on studies of type Ia supernovae).

Let us emphasize that the performed analysis crucially depends on
the accepted value of secular increase in the length of
day~$ {\dot T}_{\rm E} $.
The qualitative idea of such analysis was put forward in our
work~\cite{dum01}, and in the first quantitative study of this
subject~\cite{dum03} we used the value corrected for the ancient eclipses,
$ {\dot T}_{\rm E}  =  1.4{\cdot}10^{-5}{\,}{\rm s/yr} $,
which was considered by some researchers as the best option~\cite{ste84}.
As a result, we arrived at the substantially reduced magnitude of
the local Hubble constant,
$ H_0^{\rm (loc)}  =  33{\pm}5\,{\rm (km/s)/Mpc} $,
which had no reasonable interpretation.
On the other hand, when in the later work~\cite{dum08} we
employed~$ {\dot T}_{\rm E} $ derived purely from the set of astrometric
observations in the telescopic era~\cite{sid02} without any further
corrections, the resulting value of~$ H_0^{\rm (loc)} $ was found to be
in accordance with the large-scale cosmological data.

\bigskip

\subsection{The faint young Sun paradox}
\label{subsec:Sun}

\smallskip

Yet another appealing example for the existence of local Hubble expansion
is the problem of insufficient flux of energy from Sun to the Earth in
the past; e.g., review~\cite{fel12}.
Namely, according to the modern models of stellar evolution, the solar
luminosity increases by approximately 30\,\%~during the period after its birth
(about~$ 5{\cdot}10^9 $\,yr).
This means that the energy input to the Earth's climate system, e.g.,
2--4~billion years ago was appreciably less than now and, therefore,
the most part of water must be in a frozen state.
This would preclude the geological and biological evolution of the Earth
and contradicts a number of well-established facts on the existence of
considerable volumes of liquid water in that period of time.
Although a lot of attempts were undertaken to resolve this problem
by the inclusion of additional influences (first of all, the atmospheric
greenhouse effect), no definitive solution is available by now.

An interesting option was suggested recently by K{\v{r}}{\'i}{\v{z}}ek and
Somer~\cite{kri12,kri15}, who proposed to take into consideration
the local Hubble expansion of the Earth's orbit.
As a result, the Sun--Earth distance in the past would be appreciably
less than now and, consequently, the solar irradiation of the Earth's
surface increased.
In particular, the quantitative analysis performed in the above-cited
papers have shown that at $ H_0^{\rm (loc)}  \approx 0.5  H_0 $
expansion of the Earth's orbit compensates the increasing solar luminosity
with very good accuracy; so that the Earth's surface received almost
the same flux of energy in the past $ 3.5{\cdot}10^9 $\,yr and will
continue to do so for a~considerable period in future.

From our point of view, the above-mentioned idea is very promising.
Unfortunately, the value of local Hubble constant used in these papers
is poorly consistent with the one derived from our analysis of
the Earth--Moon system in Section~\ref{subsec:Moon},
$ H_0^{\rm (loc)}  \approx 0.85 \, H_0 $.
So, it is interesting to check if the same mechanism will work at
other rates of the local Hubble expansion?
Such analysis was performed in our recent work~\cite{dum15}.

\begin{figure}[h!]
\begin{center}
\includegraphics[width=11.9cm]{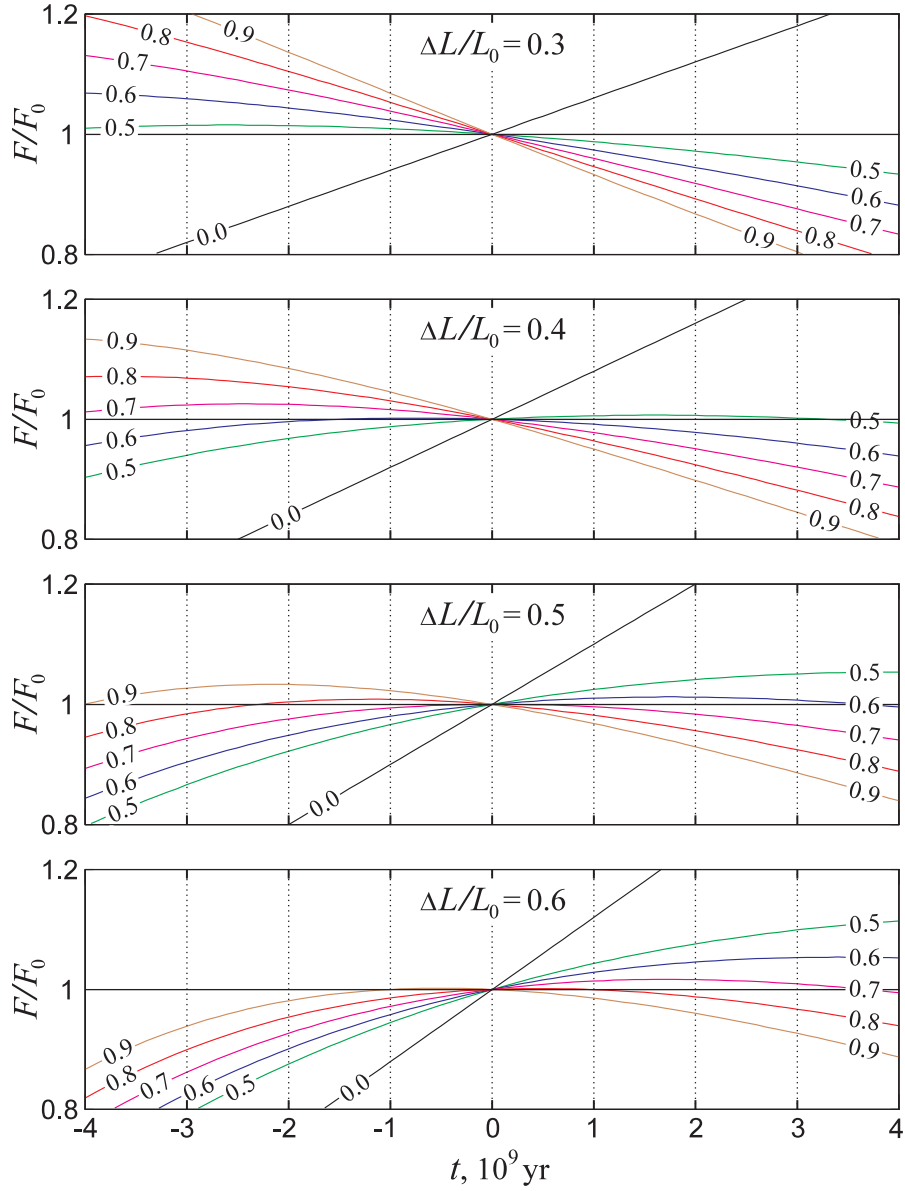}
\caption{\label{fig:Lumin-Time}
Temporal variations in solar irradiation of the Earth's surface~$ F / F_0 $
for different models of solar evolution (characterized
by~$ \Delta L / L_0 $) and various rates of the local Hubble expansion
(numbers near the curves denote the ratio~$ H_0^{\rm (loc)} \! / H_0 $).
The straight lines marked by 0.0 correspond to the case when the local
Hubble expansion is absent at all.}
\end{center}
\end{figure}

Namely, let solar luminosity increase linearly with time:

\begin{equation}
\label{eq:Lumin-time}
L(t) = L_0 + ( \Delta L / \Delta T ) \, t \, ,
\end{equation}

\smallskip

\noindent where $ L_0 $~is its present-day value (at $ t  = 0 $), and
$ \Delta L $~is the variation of luminosity over the time interval
$ \Delta T  =  5{\cdot}10^9 $\,yr
(for the sake of estimate, we shall use here the rounded values).
Then, assuming validity of the standard relation~(\ref{eq:Hubble_local_rate}),
a~temporal variation in the irradiation of the Earth's surface
can be found from a~simple geometric consideration.
The resulting curves for a number of hypothetical solar models with
$ \Delta L / L_0  =  0.3, \, 0.4, \, 0.5, \, 0.6 $ and various
rates of the local Hubble expansion
$ H_0^{\rm (loc)}  = 0.5, \, 0.6, \, 0.7, \, 0.8, \, 0.9 \, H_0 $
are presented in Figure~\ref{fig:Lumin-Time}.

It is seen that the K{\v{r}}{\'i}{\v{z}}ek--Somer case
($ \Delta L / L_0  =  0.3$, $H_0^{\rm (loc)}  / H_0  =  0.5 $)
really provides a~very stable energy input to the Earth for a~few billion
years both in the past and future.
At the higher rates of the local Hubble expansion (which would be more
consistent with our analysis of the Earth--Moon dynamics), a~quite
favorable situation exists, for example, at
$ \Delta L / L_0  = 0.5 $ and $ H_0^{\rm (loc)}  / H_0  = 0.8 $:
the solar irradiation at $ t < 0 $ is almost as stable as in
the K{\v{r}}{\'i}{\v{z}}ek--Somer case, and more appreciable variation
at $ t > 0 $ is not so important because we actually do not know
the Earth's evolution in the future.

Is it reasonable to consider the solar model with
$ \Delta L / L_0  = 0.5 \, $?
In fact, such enhanced variations~$ \Delta L $ were typical for
the first quantitative models of the Sun~\cite{sch58}.
However, the subsequent investigations resulted in the progressively
less values of~$ \Delta L $; and it is commonly accepted now that
the increase in luminosity amounts to about 7\,\%~per Gyr over the past
evolution of 4.57~Gyr.
Nevertheless, we may imagine processes like mixing in the solar interiors
to change this value.
This would imply the star with a small convective core.
The problem is that the Sun is just at the limit of mass where convective
cores appear~\cite{mae16}.

Of course, one should keep in mind that the above calculations of solar
irradiation cannot be immediately confronted with the relevant data from
paleoclimatology, because it is necessary to take into account a~lot of
additional geophysical and geochemical processes, first of all,
the greenhouse effect.
From this point of view, the Earth--Moon system discussed in
Section~\ref{subsec:Moon} represents a~more ``clean'' case,
where the probable local Hubble expansion is less obscured by
other phenomena.

\subsection{Other systems}
\label{subsec:Other}

\bigskip

A number of other effects in the Solar system that might be associated
with local Hubble expansion have been already listed in the beginning of
Section~\ref{sec:Observations}.
Unfortunately, they are much less studied than the lunar tidal catastrophe
and the faint young Sun paradox.
So, we shall not discuss them in the present article; for more details,
see papers~\cite{kri12,kri15}.

Besides, a few researchers studied dynamics of all solar-system bodies,
including the major asteroids, on the basis of data by optical and radio
astrometry collected in the last decades~\cite{pit05a,pit05b}.
Their conclusion was that, in general, a self-consistent picture of
planetary motion (the high-precision ephemerides) can be obtained
without taking into account any local cosmological influences.
However, it should be kept in mind that such analyses involved a lot
of fitting parameters, which were attributed, e.g., to the unknown
masses of asteroids, solar oblateness, effects of the solar wind
on radio wave propagation, etc.
On the other hand, \textit{the probable Hubble expansion was never
included into their equations in explicit form.}
So, the small resulting residuals might be merely a mathematical fact:
it is well known from statistics that any empirical data can be
fitted as accurately as desirable if the number of free parameters
becomes sufficiently large.

If the local Hubble expansion is present in the Solar system, it should
be naturally expected also in galaxies.
Unfortunately, the entire pattern of galaxy evolution is very complicated
by the formation of stars and their proper motions.
So, as far as we know, the problem of cosmological effects at the scale
of galaxies remains completely unexplored by now.

A much more elaborated subject is Hubble expansion in the local
intergalactic volume.
It was believed for a long time that the standard Hubble flow can be
traced only at the distances starting from 5--10\,Mpc, where it becomes
possible to introduce the average cosmological matter density.
Nevertheless, by the end of the 20th century, the Hubble flow was detected
also at the considerably less scales, down to 1--2\,Mpc.
At the same time, the concept of dark-energy-dominated Universe became
the main paradigm in cosmology.
So, it was natural to explain both the presence of the Hubble flow at
the sufficiently small scales and its regularity (``quiescence'') just
by the perfectly-uniform dark energy (or $ \Lambda $-term)~\cite{che03,ekh01}.
Unfortunately, it remains unclear by now if the effective value of
Hubble constant in the Local Group is smaller or larger than at
the global scales and, therefore, if the relation~(\ref{eq:H_0_ratio})
between $ H_0^{\rm (loc)} $~and~$ H_0 $ is applicable in this situation?

Let us mention also that the most of available theoretical works on
the dynamics of galaxies in the Local Group are based on
the effective gravitational forces derived from Kottler
metric~(\ref{eq:Kottler_metric}):

\begin{equation}
\label{eq:Eff_Force}
F_{\rm eff} (r) = M_1
  {\biggl( - \frac{G M_2}{r^2} + \frac{\Lambda \, r}{3} \biggr)} \, ;
\end{equation}


\pagebreak
\noindent the last term often being called the ``antigravity'' force.
Unfortunately, such treatment has a limited scope of applicability:
Firstly, as was already mentioned in Section~\ref{sec:Dark_energy},
the static metric~(\ref{eq:Kottler_metric}) does not possess
a correct cosmological asymptotics at infinity and, therefore,
the corresponding force~(\ref{eq:Eff_Force}) is unable to describe
the entire Hubble flow, including the large distances.
Secondly, strictly speaking, the above-written effective force is adequate
\textit{only for the restricted two-body problem}
(where $ M_1 $~is the mass of a test particle, and $ M_2 $~is the mass
of the central gravitating body).
This is evident, in particular, from the fact that masses~$ M_1 $ and~$ M_2 $ appear in expression~(\ref{eq:Eff_Force}) by different ways.
So, there is no reason to assume validity of this formula when $ M_1 $~and~$ M_2 $ are comparable to each other or, especially, to apply it to the many-body problem.

\section{Concluding remarks}
\label{sec:Conclusion}

\begin{enumerate}
\item
Despite a lot of theoretical works rejecting the possibility of local
Hubble expansion, we believe that this problem is still unresolved:
Firstly, the available arguments often contradict each other.
Secondly, the most of them become inapplicable to the case when
the Universe is dominated by the perfectly-uniform dark energy
(or~$ \Lambda $-term).
Moreover, a~self-consistent theoretical treatment of the simplest
models (such as the restricted two-body problem against
the $ \Lambda $-background) demonstrates a~principal possibility of
the local cosmological influences: the Hubble expansion is not suppressed
completely in the vicinity of a~massive body.
\item
A few long-standing problems in planetology, geophysics, and celestial
mechanics can be well resolved by the assumption of local Hubble expansion
whose rate is comparable to that at the global scales.
It is quite surprising that many theorists believe that the possibility of
local cosmological influences is strictly prohibited just by the available
observational data, while a~lot of observers believe that there are
irrefutable theoretical proofs that Hubble expansion is absent at small
scales.
\item
However, the important conceptual question still persists:
What is the spatial scale from which the cosmological expansion no longer
takes place?
This is of crucial importance since otherwise, as pictorially explained
by Misner et al.~\cite[p.\,719]{mis73}, the ``meter stick'' will also
expand and, therefore, it will be meaningless to speak about any
expansion at all...
We cannot give a definitive numerical answer to this question.
However, we believe that the systems dominated by non-gravitational
interactions should not experience the cosmological expansion
(e.g.,~the meter stick, the solid Earth, etc. do not expand).
\end{enumerate}

\pagebreak

\section*{Acknowledgements}
I am grateful to Yurij V.~Baryshev, Sergei M.~Kopeikin, Michal K{\v{r}}{\'i}{\v{z}}ek,
Andr\'e Maeder, and Marek Nowakowski for valuable discussions of the problems
outlined in this paper.

\bigskip


\begin{thebibliography}{99}

\smallskip

\providecommand{\url}[1]{\texttt{#1}}
\providecommand{\urlprefix}{URL }
\expandafter\ifx\csname urlstyle\endcsname\relax
  \providecommand{\doi}[1]{doi:\discretionary{}{}{}#1}\else
  \providecommand{\doi}{doi:\discretionary{}{}{}\begingroup
  \urlstyle{rm}\Url}\fi
\providecommand{\eprint}[2][]{\url{#2}}

\bibitem{and95}
Anderson, J.\,L.: Multiparticle dynamics in an expanding Universe.
\newblock Phys. Rev. Lett. \textbf{75} (1995), 3602.

\bibitem{bal06}
Balaguera-Antol{\'i}nez, A., B{\"o}hmer, C.\,G., and Nowakowski, M.: Scales set
  by the cosmological constant.
\newblock Class. Quant. Grav. \textbf{23} (2006), 485.

\bibitem{bon00}
Bonnor, W.\,B.: Local dynamics and the expansion of the Universe.
\newblock Gen. Rel. Grav. \textbf{32} (2000), 1005.

\bibitem{bou04}
Bourda, G. and Capitaine, N.: Precession, nutation, and space geodetic
  determination of the Earth's variable gravity field.
\newblock Astron. Astrophys. \textbf{428} (2004), 691.

\bibitem{car98}
Cardona, J.\,F. and Tejeiro, J.\,M.: Can interplanetary measures bound the
  cosmological constant?
\newblock Astrophys. J. \textbf{493} (1998), 52.

\bibitem{che03}
Chernin, A., Teerikorpi, P., and Baryshev, Y.: Why is the Hubble flow so
  quiet?
\newblock Adv. Space Res. \textbf{31} (2003), 459.

\bibitem{coo98}
Cooperstock, F.\,I., Faraoni, V., and Vollick, D.\,N.: The influence of the
  cosmological expansion on local systems.
\newblock Astrophys. J. \textbf{503} (1998), 61.

\bibitem{cox02}
Cox, C.\,M. and Chao, B.\,F.: Detection of a~large-scale mass redistribution
  in the terrestrial system since~1998.
\newblock Science \textbf{297} (2002), 831.

\bibitem{dav03}
Davis, T.\,M., Lineweaver, C.\,H., and Webb, J.\,K.: Solutions to the tethered
  galaxy problem in an expanding Universe and the observation of receding
  blueshifted objects.
\newblock Amer. J. Phys. \textbf{71} (2003), 358.

\bibitem{dic64}
Dicke, R.\,H. and Peebles, P.\,J.\,E.: Evolution of the Solar system and the
  expansion of the Universe.
\newblock Phys. Rev. Lett. \textbf{12} (1964), 435.

\bibitem{dic94}
Dickey, J.\,O. et~al.: Lunar laser ranging: A~continuing legacy of the Apollo
  program.
\newblock Science \textbf{265} (1994), 482.

\bibitem{dom01}
Dom\'{i}nguez, A. and Gaite, J.: Influence of the cosmological expansion on
  small systems.
\newblock Europhys. Lett. \textbf{55} (2001), 458.

\bibitem{dum01}
Dumin, Y.\,V.: Using the lunar laser ranging technique to measure the local
  value of Hubble constant.
\newblock Geophys. Res. Abstr. \textbf{3} (2001), 1965.

\bibitem{dum03}
Dumin, Y.\,V.: A~new application of the lunar laser retroreflectors: Searching
  for the `local' Hubble expansion.
\newblock Adv. Space Res. \textbf{31} (2003), 2461.

\bibitem{dum07}
Dumin, Y.\,V.: Comment on `Progress in lunar laser ranging tests of
  relativistic gravity'.
\newblock Phys. Rev. Lett. \textbf{98} (2007), 059\,001.

\bibitem{dum08}
Dumin, Y.\,V.: Testing the dark-energy-dominated cosmology by the solar-system
  experiments.
\newblock In: H.~Kleinert, R.~Jantzen, and R.~Ruffini (Eds.),
  \emph{Proc. 11th~Marcel Grossmann meeting on General Relativity}, p.~1752.
  World Sci., Singapore, 2008.

\bibitem{dum11}
Dumin, Y.\,V.: Perturbation of a planetary orbit by the Lambda-term (dark
  energy) in Einstein equations.
\newblock In: N.~Capitaine (Ed.), \emph{Proc. Journ{\'e}es 2010 Syst{\`e}mes de
  r{\'e}f{\'e}rence spatio-temporels: New challenges for reference systems and
  numerical standards in astronomy}, p.~276. Observ. Paris, 2011.

\bibitem{dum15}
Dumin, Y.\,V.: The faint young Sun paradox in the context of modern cosmology.
\newblock Astronomicheskii Tsirkulyar (Astron.\ Circular) \textbf{1623} (2015),
  1.
\newblock {\tt http://comet.sai.msu.ru/${}^{\sim}$gmr/AC/AC1623.pdf}.

\bibitem{ein45}
Einstein, A. and Straus, E.\,G.: The influence of the expansion of space on the
  gravitation\, fields\, surrounding\, the\, individual\, stars.\,
\newblock Rev.\ Mod.\ Phys. \textbf{17} (1945), 120.

\bibitem{ekh01}
Ekholm, T., Baryshev, Y., Teerikorpi, P., Hanski, M.\,O., and Paturel, G.:
  On the quiescence of the Hubble flow in the vicinity of the Local Group:
A~study using galaxies with distances from the Cepheid PL-relation.
\newblock Astron. Astrophys. \textbf{368} (2001), L17.

\bibitem{far07}
Faraoni, V. and Jacques, A.: Cosmological expansion and local physics.
\newblock Phys. Rev. D \textbf{76} (2007), 063\,510.

\bibitem{fel12}
Feulner, G.: The faint young Sun problem.
\newblock Rev. Geophys. \textbf{50} (2012), RG2006.

\bibitem{hac08}
Hackmann, E. and L{\"a}mmerzahl, C.: Geodesic equation in
  Schwarzschild-(anti-) de~Sitter space-times: Analytical solutions and
  applications.
\newblock Phys. Rev. D \textbf{78} (2008), 024\,035.

\bibitem{ior06}
Iorio, L.: Can solar system observations tell us something about the
  cosmological constant?
\newblock Int. J. Mod. Phys. D \textbf{15} (2006), 473.

\bibitem{kag06}
Kagramanova, V., Kunz, J., and L{\"a}mmerzahl, C.: Solar system effects in
  Schwarzschild--de~Sitter space--time.
\newblock Phys. Lett. B \textbf{634} (2006), 465.

\bibitem{kau68}
Kaula, W.: \emph{An introduction to planetary physics: The terrestrial
  planets}.
\newblock J.\ Wiley~\& Sons, New York, 1968.

\bibitem{kli05}
Klioner, S.\,A. and Soffel, M.\,H.: Refining the relativistic model for Gaia:
  Cosmological effects in the BCRS.
\newblock In: C.~Turon, K.\,S.~O'Flaherty, and M.\,A.\,C.~Perryman (Eds.),
  \emph{Proc. Symp. The three-dimensional Universe with Gaia (ESA SP-576)},
  p.~305. ESA Publ. Division, Noordwijk, Netherlands, 2005.

\bibitem{kot18}
Kottler, F.: Uber die physikalischen Grundlagen der Einsteinschen
  Gravitationstheorie.
\newblock Ann. Phys. (Leipzig) \textbf{56} (1918), 401.

\bibitem{kra80}
Kramer, D., Stephani, H., MacCallum, M., and Herlt, E.: \emph{Exact solutions
  of Einstein's field equations}.
\newblock Cambridge University Press, Cambridge, 1980.

\bibitem{kri12}
K{\v{r}}{\'i}{\v{z}}ek, M.: Dark\, energy\, and\, the\, anthropic\,
  principle. \,
\newblock New\, Astron.\, \textbf{17} (2012), 1.

\bibitem{kri15}
K{\v{r}}{\'i}{\v{z}}ek, M. and Somer, L.: Manifestations of dark energy in the
 Solar system.
\newblock Grav. Cosmol. \textbf{21} (2015), 59.

\bibitem{lan76}
Landau, L.\,D. and Lifshitz, E.\,M.: \emph{Mechanics}.
\newblock Pergamon Press, Oxford, 1976, 3rd~edn.

\bibitem{lin05}
Lineweaver, C.\,H. and Davis, T.\,M.: Misconceptions about the Big Bang.
\newblock Sci. American \textbf{292, no.~3} (2005), 36.

\bibitem{mae16}
Maeder, A.: Private communication  (2016).

\bibitem{mvi33}
McVittie, G.\,C.: The mass-particle in an expanding Universe.
\newblock Mon. Not. Royal Astron. Soc. \textbf{93} (1933), 325.

\bibitem{mis73}
Misner, C.\,W., Thorne, K.\,S., and Wheeler, J.\,A.: \emph{Gravitation}.
\newblock W.\,H.~Freeman~\& Co., San Francisco, 1973.

\bibitem{noe71}
Noerdlinger, P.\,D. and Petrosian, V.: The effect of cosmological expansion on
  self-gravitating ensembles of particles.
\newblock Astrophys. J. \textbf{168} (1971), 1.

\bibitem{pit05a}
Pitjeva, E.\,V.: High-precision ephemerides of planets--EPM and determination
  of some astronomical constants.
\newblock Solar System Res. \textbf{39} (2005), 176.

\bibitem{pit05b}
Pitjeva, E.\,V.: Relativistic effects and solar oblateness from radar
  observations of planets and spacecraft.
\newblock Astron. Lett. \textbf{31} (2005), 340.

\bibitem{sch58}
Schwarzschild, M.: \emph{Structure and evolution of the stars}.
\newblock Princeton Univ.\ Press, Princeton, N.J., 1958.

\bibitem{ser07}
Sereno, M. and Jetzer, P.: Evolution of gravitational orbits in the expanding
  Universe.
\newblock Phys. Rev. D \textbf{75} (2007), 064\,031.

\bibitem{sid02}
Sidorenkov, N.\,S.: \emph{Physics of the Earth's rotation instabilities}.
\newblock Nauka-Fizmatlit, Moscow, 2002.
\newblock In Russian.

\bibitem{ste84}
Stephenson, F.\,R. and Morrison, L.\,V.: Long-term changes in the rotation of
  the Earth: 700~{B.C.} to {A.D.}~1980.
\newblock Phil. Trans. Royal Soc. Lond. A \textbf{313} (1984), 47.

\bibitem{yod83}
Yoder, C.\,F. et~al.: Secular variation of Earth's gravitational
  harmonic~{$ J_2 $} coefficient from Lageos and nontidal acceleration of
  Earth rotation.
\newblock Nature \textbf{303} (1983), 757.

\end{thebibliography}

\end{document}